\documentclass{article}
\usepackage{netcod09}

\usepackage{graphicx} 
\usepackage{color}
\usepackage{psfrag}
\usepackage{epsfig}

\usepackage{booktabs} 
\usepackage{array} 
\usepackage{paralist} 
\usepackage{verbatim} 
\usepackage{amsmath,amssymb,amsthm}

\newtheorem{lemma}{Lemma}

\newcommand{\E}{{\rm E}}

\newcommand{\ul}[1]{\underline{#1}}

\newcommand{\set}[1]{{\cal #1}}

\begin{document}
\topmargin = 0mm

\itwtitle{Pipelined Encoding for Deterministic \\ and Noisy Relay Networks}

\itwauthor{Gerhard Kramer}
{Department of Electrical Engineering\\
University of Southern California \\
Los Angeles, CA, USA\\
Email: gkramer@usc.edu}

\itwmaketitle



\begin{itwabstract}
Recent coding strategies for deterministic and noisy relay networks are related to the pipelining of block Markov encoding. For deterministic networks, it is shown that pipelined encoding improves encoding delay, as opposed to end-to-end delay. For noisy networks, it is observed that decode-and-forward exhibits good rate scaling when the signal-to-noise ratio (SNR) increases.
\end{itwabstract}


\begin{itwpaper}
\itwsection{Introduction}

Consider a network represented by a graph $\set{G}=(\set{V},\set{E})$ where $\set{V}$ is a set of vertices (or nodes) and $\set{E}$ is a set of directed edges. There are $M$ messages $W_m$, $m=1,2,\ldots,M$, and every message is associated with one of the nodes. As described in~\cite[Ch.~3]{Kramer-now-cc:06}, with every node $u$ we further associate one channel input $X_u$ and one channel output $Y_u$. The output $Y_v$ is a (generally noisy) function of the channel inputs $X_u$ of those nodes $u$ having directed edges $(u,v)\in\set{E}$. A central clock governs the operation of the network~\cite{Ratnakar:06}. The clock ticks $n$ times and node $u$ is permitted to transmit symbol $X_u^{(i)}$ after clock tick $i-1$ and before clock tick $i$, $i=1,2,\ldots,n$. The symbol $Y_u^{(i)}$ appears at clock tick $i$. The network is also {\it causal} in the sense that $X_u^{(i)}$ is a function of messages at node $u$ and the past outputs $Y_u^{i-1}=Y_u^{(1)},Y_u^{(2)},\ldots,Y_u^{(i-1)}$. 
This graphical model was considered in~\cite{KramerSavari:06}, for example, where edge-cut bounds were developed.

Suppose there is one message only. The paper~\cite{Avestimehr:08} develops achievable rates by using a compress-and-forward (CF) strategy.  Suppose further that the channels are deterministic, which means that $Y_v$ is a function of $\{X_u: (u,v)\in\set{E}\}$. The paper \cite{Avestimehr:07} develops interesting achievable rates. A simpler version of this problem with broadcasting and without interference was considered in~\cite{Aref:80,Ratnakar:06} where capacity theorems were discovered. An even more basic model was considered in~\cite{Ahlswede:00} where there is no broadcasting and no interference. Broadcast erasure and finite-field networks are considered in~\cite{LunAllerton04,Dana:06,Smith:07,Bhadra:06,Avestimehr:07}.

One goal of this document is to revisit and clarify the coding methodology and analysis of~\cite{Ratnakar:06,Avestimehr:07}. A second goal is to point out relations to block-Markov coding and decoding methods~\cite{Cover79,XieK:04,KGG:05}. A third goal is to state the fact that decode-and-forward (DF) exhibits good signal-to-noise ratio (SNR) scaling because it removes interference~\cite{XieK:04}.

\itwsection{Cuts and Bounds}
\label{sec:cuts}

Consider a set $\set{S}$ of nodes and let $R$ be the rate of the message. Let $\Lambda$ be the collection of all cuts $(\set{S},\set{S}^c)$ that separate $s$ from one of the destinations, where $\set{S}^c$ is the complement of $\set{S}$ in $\set{V}$. A standard cut-set bound~(see~\cite[Ch.~14]{CoverThomas} or~\cite[Sec.~10.2]{Kramer-now-muit:07}) specifies that
reliable communication requires
\begin{align}
  R & \le \max_{P_{X_1X_2\cdots X_{|\set{V}|}}(\cdot)} \:
       \min_{(\set{S},\set{S}^c) \in \Lambda} \:
       I( X_{\set{S}} ; Y_{\set{S}^c} | X_{\set{S}^c} )
    \label{eq:converse2}
\end{align}
where $X_\set{S}=\{X_u: u\in\set{S}\}$ and similarly for $Y_{\set{S}^c}$.

A simplification of \eqref{eq:converse2} is achieved by defining the two {\it boundaries} of $\set{S}$ as
\begin{align}
  & \beta_1(\set{S})=\{u: (u,v)\in(\set{S},\set{S}^c) \} \label{eq:beta1} \\
  & \beta_2(\set{S})=\{v: (u,v)\in(\set{S},\set{S}^c) \}. \label{eq:beta2}
\end{align}
Let $\set{A}-\set{B}=\{a: a\in\set{A},a\notin\set{B}\}$ and observe that
$X_{\set{S}}-X_{\set{S}^c}Y_{\beta_2(\set{S})}-Y_{(\set{S}^c-\beta_2(\set{S}))}$ forms a Markov chain. We thus have
\begin{align}
  I( X_{\set{S}} ; Y_{\set{S}^c} | X_{\set{S}^c} )
  & = I(X_{\set{S}} ; Y_{\beta_2(\set{S})} Y_{\set{S}^c-\beta_2(\set{S})}  | X_{\set{S}^c} ) \nonumber \\
  & = I(X_{\set{S}} ; Y_{\beta_2(\set{S})} | X_{\set{S}^c} ).
  \label{eq:converse3}
\end{align}
For deterministic networks, \eqref{eq:converse3} becomes
\begin{align}
  I(X_{\set{S}} ; Y_{\beta_2(\set{S})} | X_{\set{S}^c} )
  & = H( Y_{\beta_2(\set{S})} | X_{\set{S}^c} ).
  \label{eq:converse3a}
\end{align}
For deterministic networks with broadcasting but no interference, or Aref networks, the channel output of node $v$ is a vector $Y_v=[Y_{u,v}: (u,v)\in\set{E}]$ where $Y_{u,v}=f_{u,v}(X_u)$ for some function $f_{u,v}(\cdot)$. The point is that node $v$ experiences no interference, a situation encountered if the transmitters use frequency or time-division multiplexing (FDM/TDM). We simplify the expression \eqref{eq:converse3a} as follows:
\begin{align}
  H( Y_{\beta_2(\set{S})} | X_{\set{S}^c} )
  & \le H(Y_{\beta_2(\set{S})})  \nonumber \\
  & \le \sum_{u\in\beta_1(\set{S})} H(Y_{u,\beta_2(\set{S})})  
  \label{eq:converse3b}
\end{align}
where the two inequalities hold with equality if the $X_u$, $u\in\set{V}$, are statistically independent. It turns out that independent $X_u$ are best for Aref networks (see~\cite[Lemma~1]{Ratnakar:06}).

Summarizing, the cut-set bound is
\begin{align}
  R & \le \max_{P_{X_1X_2\cdots X_{|\set{V}|}}(\cdot)} \:
       \min_{(\set{S},\set{S}^c) \in \Lambda} \:
       \mathrm{Value}(\set{S},\set{S}^c)
    \label{eq:converse4}
\end{align}
where the {\it value} of the cut $(\set{S},\set{S}^c)$ is
\begin{align}
    & \mathrm{Value}(\set{S},\set{S}^c) \nonumber \\
    & = \left\{ \begin{array}{ll}
    I(X_{\set{S}} ; Y_{\beta_2(\set{S})} | X_{\set{S}^c} ) & \text{in general} \\
    H( Y_{\beta_2(\set{S})} | X_{\set{S}^c} ) & \text{for deterministic networks} \\
    \sum_{u\in\beta_1(\set{S})} H(Y_{u,\beta_2(\set{S})})  & \text{for Aref networks}
    \end{array} \right.
    \label{eq:converse5}
\end{align}
and for Aref networks the optimization over joint input distributions results in a product distribution.

For example, consider the Aref network in Fig.~\ref{fig:standardExample} and $\set{S}=\{1,2,3,7\}$ so that $Y_{\set{S}^c}=\{Y_4,Y_5,Y_6\}$ where $Y_4=[Y_{2,4},Y_{3,4}]$, $Y_5=Y_{4,5}$, and $Y_6=[Y_{2,6},Y_{5,6}]$.
We have $\beta_1(\set{S})=\{2,3\}$, $\beta_2(\set{S})=\{4,6\}$, and
\begin{align}
  \mathrm{Value}(\set{S},\set{S}^c) = H(Y_{2,4} Y_{2,6}) + H(Y_{3,4}).
  \label{eq:converse5a}
\end{align}
Observe that we must consider the joint entropy of $Y_{2,4}$ and $Y_{2,6}$, and separately the marginal entropy of $Y_{3,4}$. This separation occurs because the inputs $X_u$, $u\in\set{V}$, are statistically independent.

\begin{figure}[t]
  \begin{center}
    \includegraphics[scale=.55]{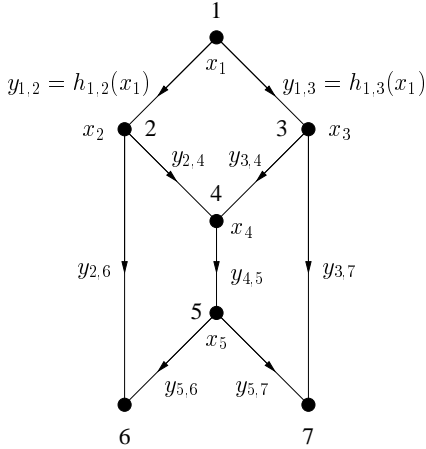}
  \end{center}
  \caption{Example of a deterministic relay network with no interference.}
  \label{fig:standardExample}
\end{figure}

\itwsection{Multicast Coding}
\label{sec:multicast-coding-aref}
%
\noindent
{\bf Aref Networks} \smallskip

We begin with Aref networks and consider acyclic directed graphs.  Suppose we use every edge $(u,v)$ exactly $n$ times by activating the nodes in topological order. For example, in Fig.~\ref{fig:standardExample} we
activate node~1 for $n$ clock ticks, then we activate node~2 for $n$ clock ticks, then node~3, and so forth. Observe that every node buffers its received symbols so that every transmit vector is a function of one message sent by the source node. We thus {\it pipeline} transmission to achieve a continuous transmission rate that is the same as the individual-activation rate. This block structure was also used in~\cite{Ahlswede:00} and it reminds us of the {\it block Markov} coding structure of~\cite{Cover79} except that pipelining requires no Markov dependencies. We shall return to this issue below when we consider interference.

We continue with our achievability proof, which is the same as in~\cite{Ratnakar:06} with minor differences. The proof in~\cite{Ratnakar:06}, in turn, follows the steps of~\cite[Sec.~V.A]{Ahlswede:00} with the main difference being the use of typical sequences. The reason for repeating the proof here is to later point out subtle issues for deterministic and noisy networks. We use the same typical sequence sets $T_{\delta}^n(P_{X_u})$ and $T_{\delta}^n(P_{Y_u})$ as in~\cite{Ratnakar:06}.  Let $f_u^n(\cdot)=[f_v^{(i)}(\cdot): i=1,2,\ldots,n]$.

\smallskip
{\bf Codebooks}. Choose $P_{X_1}(\cdot),P_{X_2}(\cdot),\ldots,
P_{X_{|\set{V}|}}(\cdot)$ and suppose that the message is at node $s=1$. At node 1, choose
$f_1^n(\cdot)$ to map each of the indices in $\{1,2,\ldots,2^{nR}\}$
to a sequence $\ul{x}_1$ drawn uniformly from
$T^n_{\delta}(P_{X_1})$.  At node $u$, $u\ne1$, choose $f_u^n(\cdot)$
to map each sequence in $T^n_{\delta}(P_{Y_u})$ to a
sequence drawn uniformly from $T^n_{\delta}(P_{X_u})$. Note that we
have $\ul{y}_u\in T^n_{\delta}(P_{Y_u})$ for all
$u$ since $\ul{x}_u\in T^n_{\delta}(P_{X_u})$
(see~\cite[Lemma~4]{Ratnakar:06}).

\smallskip
{\bf Encoding}. Node $u=1$ transmits
$\ul{x}_1(w)=f_1^n(w)$. Node $u$, $u\ne1$, transmits
$\ul{x}_u(w)=f_u^n\left(\ul{y}_u(w)\right)$. Note that we have labeled $\ul{y}_u$
with the message $w$. This makes sense for deterministic networks because $w$ is mapped to
unique $\ul{x}_{\set{V}}$ and $\ul{y}_{\set{V}}$ once the code books are chosen.

\smallskip
{\bf Decoding}. Destination node $t$ puts out
\begin{align}
   \hat{w}_t\left(\ul{y}_t(w)\right)
   = \begin{cases} \mbox{error}
     & \text{if $\ul{y}_t(w')
                =\ul{y}_t(w)$ for $w' \neq w$} \\
     w & \text{otherwise.}
   \end{cases}
   \label{eq:decoder}
\end{align}

\smallskip
{\bf Analysis}. We say that node $u$ {\it can distinguish} between $w$ and $w'$ if
\begin{align}
   \ul{y}_u(w) \ne
   \ul{y}_u(w').
   \label{eq:yne}
\end{align}
Let  $\set{S}(w,w')$ be the set of nodes that can distinguish $w$ and $w'$ and observe that the event $\set{S}(w,w') = \set{S}$ is simply the event
\begin{align}
   \{ \ul{Y}_{\set{S}}(w) \ne
   \ul{Y}_{\set{S}}(w') \} \cap
   \{\ul{Y}_{\set{S}^c}(w) =
   \ul{Y}_{\set{S}^c}(w')\}.
\label{eq:yeq}
\end{align}
We may as well consider $s\in\set{S}(w,w')$. An error occurs at destination node $t$  if $t\notin\set{S}(w,w')$, i.e., if $(\set{S}(w,w'),\set{S}^c(w,w'))$ is a cut between nodes $s$ and $t$. Let ${\Lambda}_t$ be the set of such cuts,
i.e., we define ${\Lambda}_t = \{\set{S}\subset\set{V}: s\in\set{S}, t\in\set{S}^c\}$. 

Let $\overline{P}_e(t,w,w')$ be the average probability that node $t$ cannot distinguish between $w$ and
$w'$, where the average is over the ensemble of encoding functions. We can write
\begin{align}
   \overline{P}_e(t,w,w') & = \mathrm{Pr} \left[
      \bigcup_{\set{S}\in{\Lambda}_t} \left\{
      \set{S}(w,w')=\set{S} \right\} \right] \nonumber \\
   & = \sum_{\set{S}\in{\Lambda}_t}
     \mathrm{Pr} \left[ \set{S}(w,w')=\set{S} \right].
   \label{eq:PeBound1}
\end{align}
Using \eqref{eq:yeq}, we can further write\footnote{Note that~\cite{Ratnakar:06} should have
included $\{\ul{Y}_\set{S}(w) \ne \ul{Y}_\set{S}(w')\}$ in the conditioning of its
equation (14), since the inclusion of this set is required for the conditional statistical
independence of the $\ul{X}_u(w)$ across $u$ and $w$. The text in~\cite{Ratnakar:06} is corrected by including $\{\ul{Y}_\set{S}(w) \ne \ul{Y}_\set{S}(w')\}$ in the conditioning in (14) and (19); the remaining steps are the same as in~\cite{Ratnakar:06}.}
\begin{align}
  & \mathrm{Pr} \left[ \set{S}(w,w')=\set{S} \right] \nonumber \\
  & \le 
      \mathrm{Pr} \left[ \left.
      \ul{Y}_{\set{S}^c}(w) = \ul{Y}_{\set{S}^c}(w')
      \right|
      \ul{Y}_{\set{S}}(w) \ne \ul{Y}_{\set{S}}(w')
       \right] \nonumber \\
  & = \mathrm{Pr} \left[ \left.
      \ul{Y}_{\beta_2(\set{S})}(w) = \ul{Y}_{\beta_2(\set{S})}(w')
      \right|
      \ul{Y}_{\set{S}}(w) \ne \ul{Y}_{\set{S}}(w')
       \right] \label{eq:step1} \\
  & = \mathrm{Pr} \left[ \left.
      \ul{Y}_{\beta_1(\set{S}),\beta_2(\set{S})}(w) = \ul{Y}_{\beta_1(\set{S}),\beta_2(\set{S})}(w')
      \right| \right. \nonumber \\
  & \hspace{3cm} \left.  \ul{Y}_{\beta_1(\set{S})}(w) \ne \ul{Y}_{\beta_1(\set{S})}(w')
       \right] \nonumber \\
  & = \prod_{u\in\beta_1(\set{S})}
      \mathrm{Pr} \left[ \left. 
      \ul{Y}_{u,\beta_2(\set{S})}(w)
                          = \ul{Y}_{u,\beta_2(\set{S})}(w')
      \right| \right. \nonumber \\
  & \hspace{3cm} \left. \ul{Y}_u(w)
              \ne \ul{Y}_u(w') 
      \right]
   \label{eq:ensembleindep}
\end{align}
where the last step follows because the pairs
$(\ul{X}_u(w),\ul{X}_u(w'))$,
$u\in\set{\beta}_1(\set{S})$, are statistically independent if
$\{\ul{Y}_\set{S}(w) \ne \ul{Y}_\set{S}(w')\}$ occurs, and because there is no interference.

We proceed to bound the probability in (\ref{eq:ensembleindep}). We
have $(\ul{x}_{u}(w'),\ul{y}_{u,\beta_2(\set{S})}(w'))\in
T^n_{\delta}(P_{X_uY_{u,\beta_2(\set{S})}})$ by \cite[Lemma~4]{Ratnakar:06}. The
event (\ref{eq:yeq}) thus implies
\begin{align}
  \left( \ul{X}_{u}(w'),
  \ul{Y}_{u,\beta_2(\set{S})}(w) \right)
  \in T^n_{\delta}(P_{X_u Y_{u,\beta_2(\set{S})}}).
  \label{eq:event}
\end{align}
But note that $\ul{X}_{u}(w')$ is independent of
$\ul{X}_{u}(w)$, and hence $\ul{Y}_{u,\beta_2(\set{S})}(w)$,
when conditioned on
$\{\ul{Y}_u(w)\ne\ul{Y}_u(w')\}$. The
probability of (\ref{eq:event}) occurring is thus
\begin{align}
   \left. \left| T^n_{\delta} ( P_{X_u Y_{u,\beta_2(\set{S})}} | \ 
   \ul{y}_{u,\beta_2(\set{S})}(w) ) \right|
   \right/ \left| T^n_{\delta}(P_{X_u}) \right|.
   \label{eq:div}
\end{align}
We use~\cite[Lemma~2]{Ratnakar:06} and~\cite[Lemma~3]{Ratnakar:06} to bound
\begin{align}
   & |T^n_{\delta}(P_{X_u})|
     \ge (1-\epsilon_{\delta}(n)) \cdot
         2^{n (1-\delta) H(X_u)}
   \label{eq:fancy1} \\
   & |T^n_{\delta}(P_{X_u Y_{u,\beta_2(\set{S})}} |\ul{y}_{u,\beta_2(\set{S})}(w))|
     \le 2^{n (1+\delta) H\left(X_u|Y_{u,\beta_2(\set{S})}\right)}
   \label{eq:fancy2}
\end{align}
where $\epsilon_{\delta}(n)\rightarrow0$ as $n\rightarrow\infty$.
The remaining steps are the same as in~\cite{Ratnakar:06} and we will
not repeat them here. We find that the average error probability can be made small
if $n$ is large and
\begin{align}
  R & < \min_{(\set{S},\set{S}^c)\in\Lambda} \mathrm{Value}(\set{S},\set{S}^c).
  \label{eq:probineq5}
\end{align}
Finally, we optimize over all input distributions. The result is that we can make the
overall rate approach the right-hand side of (\ref{eq:converse4})
while at the same time ensuring reliable communication. The multicast capacity of Aref networks with cycles can be
similarly achieved by constructing a time-parameterized acyclic graph
as described in~\cite[p.~146]{FF:62} or \cite{Ahlswede:00}, for example.

\bigskip \noindent
{\bf Layered Deterministic Networks} \smallskip

Relay coding for networks with interference was
considered in several recent papers~\cite{Smith:07itw,Traskov:07,Nazer:07,Avestimehr:07,Nam:08}).
However, at the moment the problem seems too difficult to solve even for 
networks with 4 nodes (the 3-node problem was solved in~\cite{GamalA:82}).
Instead, the authors of~\cite{Avestimehr:07} developed an achievable rate where
the channel inputs $X_u$, $u\in\set{V}$, are independent. Two motivations for
doing this are (1) the theory is simplified and (2) independent inputs will give
the proper capacity scaling with SNR since beamforming will not provide scaling gains
(see~\cite{Avestimehr:08}).

The coding methodology of~\cite{Avestimehr:07} uses the same random coding
and mapping at the relays as above. Furthermore, for so-called layered networks,
the encoding at the source is also the same as in~\cite{Ratnakar:06} because pipelining can be used.
The difference to~\cite{Ratnakar:06} lies in the analysis that we now outline with slight
modifications.

To begin, we add a technical step and restrict attention to messages $w$ for which
$\ul{x}_\set{V}(w) \in T^n_{\delta}(P_{X_\set{V}})$.
This step hardly reduces the rate since the code words are chosen independently via the product
distribution $P_{X_{\set{V}}}$.
We continue to use the definition that node $u$ can distinguish between $w$ and $w'$ if \eqref{eq:yne}
is true. Since the network is deterministic, every node knows
$x_\set{V}(w)$ and $x_\set{V}(w')$ and so, given $\ul{y}_u(w)$, node $u$ can check whether
\begin{align}
   (\ul{x}_\set{V}(w'),\ul{y}_u(w)) \in T^n_{\delta}(P_{X_\set{V}Y_u}).
   \label{eq:yne2}
\end{align}
Let  $\set{S}(w,w')$ be the set of nodes that can distinguish between $w$ and $w'$, i.e.,
\eqref{eq:yne2} does not occur. We note two interesting facts for deterministic networks:
\begin{itemize} 
\item the marginal typicality \eqref{eq:yne2} over $u\in\beta_2(\set{S})$ implies the joint typicality 
\begin{align}
   (\ul{x}_\set{V}(w'),\ul{y}_{\beta_2(\set{S})}(w)) \in T^n_{\delta}(P_{X_\set{V}Y_{\beta_2(\set{S})}})
   \label{eq:yne3}
\end{align}
\item the typicality \eqref{eq:yne3} implies
$\ul{y}_{\beta_2(\set{S})}(w)=\ul{y}_{\beta_2(\set{S})}(w')$ and therefore
$\ul{x}_{\set{S}^c}(w')=\ul{x}_{\set{S}^c}(w)$.
\end{itemize}
Both of the above facts are simple consequences of the definition of typical sequences (see~\cite[Lemma~4]{Ratnakar:06}).
We thus have the following result.
\begin{lemma}\label{lemma:one}
Suppose that $\ul{X}_\set{V}(w) \in T^n_{\delta}(P_{X_\set{V}})$ for all $w$.
The event $\set{S}(w,w')=\set{S}$ in \eqref{eq:yeq} then implies the event
\begin{align}
   \left( \ul{X}_\set{S}(w'),\ul{Y}_{\beta_2(\set{S})}(w),\ul{X}_{\set{S}^c}(w) \right)
   \in T^n_{\delta}( P_{X_\set{S}Y_{\beta_2(\set{S})}X_{\set{S}^c}} )
   \label{eq:event2}
\end{align}
where $X_\set{S}(w')$ is independent of $X_\set{V}(w) Y_{\set{V}}(w)$.
\end{lemma}
Lemma~\ref{lemma:one} and similar steps as \eqref{eq:div}-\eqref{eq:fancy2} give
\begin{align}
     \mathrm{Pr} \left[ \set{S}(w,w')=\set{S} \right]
     & \le 2^{-n[I(X_\set{S};Y_{\beta_2(\set{S})} X_{\set{S}^c}) - 3 \delta H(X_{\set{V}})]} \nonumber \\
     & = 2^{-n[H(Y_{\beta_2(\set{S})} | X_{\set{S}^c}) - 3 \delta H(X_{\set{V}})]}
\end{align}
Continuing as for Aref networks, we find that $R$ satisfying \eqref{eq:probineq5} is achievable, where
$\mathrm{Value}(\set{S},\set{S}^c)$ is defined in \eqref{eq:converse5} and where the $X_u$, $u\in\set{V}$,
are independent.

\bigskip \noindent
{\bf Acyclic Deterministic Networks} \smallskip

Consider next acyclic networks. We interpret the coding described in~\cite[Sec.~VI]{Avestimehr:07} as follows. Let $L$ be the length of the longest path from the source node to any destination node. Transmission is divided into $B+L-1$ length-$n$ blocks of symbols, where $B$ is a large integer, and in every block a different random code is chosen for every node. The random codes for the source node have $2^{nBR}$ code words for every block. In block $b$, $b=1,2,\ldots,B+L-1$, the source node maps the long message $w$ with $nBR$ bits to the codewords of the $b$th code.

For example, consider the network in Fig.~\ref{fig:fournodes} where nodes 1 and 4 are the message and destination nodes, respectively. We have $L=3$ and the encoding for $B=3$ is depicted in Table~\ref{table:one}. We have labeled every code word $\ul{x}_u^{(b)}$ of node $u$ in block $b$ with the channel output $\ul{y}_u^{(b-1)}$ and message of which it is a function. After the $B+L-1=5$ transmission blocks are completed, decoding can proceed by using one's favorite (ML, typicality, etc.) decoding method over all blocks of outputs. We remark that this method might be considered a special type of block Markov coding method~\cite{Cover79,XieK:04,KGG:05} with Markov dependencies across all blocks. 

\begin{figure}[t]
  \begin{center}
    \includegraphics[scale=.55]{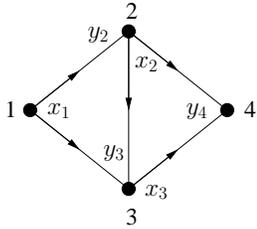}
  \end{center}
  \caption{Example of an acyclic deterministic network.}
  \label{fig:fournodes}
\end{figure}

\begin{table*}
\begin{center}
\caption{A coding strategy for the network of Fig.~\ref{fig:fournodes} for $B=3$.}
\begin{tabular}[tbh]{|c|c|c|c|c|} \hline
Block $b$ & Message & 1 Transmits & 2 Transmits & 3 Transmits \\\hline
1& $w$ & $\ul{x}_1^{(1)}(w)$ & $\cdot$ & $\cdot$ \\\hline
2& $w$ & $\ul{x}_1^{(2)}(w)$ & $\ul{x}_2^{(2)}(\ul{y}_2^{(1)}(w))$ & $\ul{x}_3^{(2)}(\ul{y}_3^{(1)}(w))$  \\\hline
3& $w$ & $\ul{x}_1^{(3)}(w)$ & $\ul{x}_2^{(3)}(\ul{y}_2^{(2)}(w))$ & $\ul{x}_3^{(3)}(\ul{y}_3^{(2)}(w))$ \\\hline
4& $\cdot$ & $\cdot$ & $\ul{x}_2^{(4)}(\ul{y}_2^{(3)}(w))$ & $\ul{x}_3^{(4)}(\ul{y}_3^{(3)}(w))$  \\\hline
5& $\cdot$ & $\cdot$ & $\cdot$ & $\ul{x}_3^{(5)}(\ul{y}_3^{(4)}(w))$  \\\hline
\end{tabular}\label{table:one}
\end{center}
\end{table*}

We wish to understand if one can improve the end-to-end (encoding and decoding) delay. Suppose we use the same pipelined encoding method as for Aref networks or layered networks. In other words, we split the message $w$ into $B$ blocks $w_1,w_2,\ldots,w_B$ each having $nR$ bits. In block $b$, the source encoder maps the message $w_b$ to its codeword $\ul{x}_1^{(b)}(w_b)$. The relays operate as before. However, note that the relay nodes experience interference, i.e., every relay node's transmission is affected by several messages.
As before, transmission is done using $B+L-1$ length-$n$ blocks of symbols, and in every block a new random code is chosen for every node.

For example, consider again the network in Fig.~\ref{fig:fournodes}. Suppose we use the ``natural" encoding depicted in Table~\ref{table:two} where we have labeled every code word $\ul{x}_u^{(b)}$ with the channel output $\ul{y}_u^{(b-1)}$ and the messages that affect them. The destination could wait until all $B+L-1=5$ blocks are received and then perform a joint decoding of all messages. As a result, we recover the rate of the strategy in Table~\ref{table:one} but with a smaller encoding delay and complexity. This might be important, for instance, if $w_1$ must be encoded before the messages $w_2$ and $w_3$ arrive at the source node. Alternatively, we could use backward decoding with a sliding window of length two. For example, by considering its outputs from blocks $b=4,5$ the destination can decode $w_3$ with the desired mutual information of $I(X_2X_3;Y_4)$, and similarly for $w_2$ and $w_1$. 

On the other hand, although the encoding delay is reduced as compared to Table~\ref{table:one}, the maximum end-to-end delay has not changed. Moreover, node 4 cannot use a forward sliding window decoder to reduce the maximum delay. For instance, consider $w_1$ which one can hope to decode after block $b=3$. However, the interference from $w_2$ in $\ul{x}_3^{(3)}(w_1,w_2)$ prevents the method from working as desired. We have also tried other encoding methods but have so far failed to reduce the end-to-end delay for general acyclic deterministic networks.

\begin{table*}
\begin{center}
\caption{A pipelining strategy for the network of Fig.~\ref{fig:fournodes} for $B=3$.}
\begin{tabular}[tbh]{|c|c|c|c|l|} \hline
Block $b$ & Message & 1 Transmits & 2 Transmits & 3 Transmits \\\hline
1& $w_1$ & $\ul{x}_1^{(1)}(w_1)$ & $\cdot$ & $\cdot$ \\\hline
2& $w_2$ & $\ul{x}_1^{(2)}(w_2)$ & $\ul{x}_2^{(2)}(\ul{y}_2^{(1)}(w_1))$ & $\ul{x}_3^{(2)}(\ul{y}_3^{(1)}(w_1))$  \\\hline
3& $w_3$ & $\ul{x}_1^{(3)}(w_3)$ & $\ul{x}_2^{(3)}(\ul{y}_2^{(2)}(w_2))$ & $\ul{x}_3^{(3)}(\ul{y}_3^{(2)}(w_1,w_2))$ \\\hline
4& $\cdot$ & $\cdot$ & $\ul{x}_2^{(4)}(\ul{y}_2^{(3)}(w_3))$ & $\ul{x}_3^{(4)}(\ul{y}_3^{(3)}(w_2,w_3))$ \\\hline
5& $\cdot$ & $\cdot$ & $\cdot$ & $\ul{x}_3^{(5)}(\ul{y}_3^{(4)}(w_3))$ \\\hline
\end{tabular}\label{table:two}
\end{center}
\end{table*}

\itwsection{SNR Scaling}
\label{sec:scaling}

A specialized SNR scaling result was developed for noisy networks in~\cite{Avestimehr:08}. The model in this paper specifies a channel gain $a^{b_{u,v}}$ for edge $(u,v)$, where $b_{u,v}$ is a positive integer. The parameter $a$ is then made large. Instead, suppose that the channel inputs $X_u$ are complex numbers and the channel outputs are
\begin{align}
   Y_v = Z_v + \sum_{u: (u,v)\in\set{E}} \sqrt{g_{u,v}} \: X_u
\end{align}
where $g_{u,v}$ is a real and positive gain coefficient, and $Z_v$ is complex Gaussian noise with independent real and imaginary parts each having variance $N/2$. The $Z_v$, $v\in\set{V}$, are independent and we add the constraint $\E[|X_u|^2] \le P$ for all $u$.

Consider the network graph. The cut-set bound \eqref{eq:converse4} is positive only if there is a Steiner tree rooted at the source node with leaves at every destination node that has non-zero gains along every edge of the tree. We use DF with block Markov encoding and sliding window decoding~\cite{XieK:04,KGG:05} along this tree, with common-message broadcasting at forks in the tree (recall that we have full-duplex nodes). This DF strategy effectively removes interference~\cite{XieK:04,KGG:05} and can achieve at least the rate
\begin{align}
   R = \log(1+g_{min}P/N)
   \label{eq:DFrate-LB}
\end{align}
where $g_{min}=\min_{u,v} g_{u,v}$.  On the other hand, the cut bound \eqref{eq:converse4} for the cut $\set{S}=s$ is at least as restrictive as
\begin{align}
   R \le \log(1+g_{max} (|\set{V}|-1)P/N)
   \label{eq:DFrate-UB}
\end{align}
where $g_{max}=\max_{u,v} g_{u,v}$ and the factor $|\set{V}|-1$ assumes that $X_s$ can be received by all other nodes. Hence, we find that at high SNR DF achieves within
\begin{align}
   & \log_2(1+g_{max} (|\set{V}|-1)P/N) - \log_2(1+g_{min}P/N) \nonumber\\
   & \approx \log_2\left( \frac{g_{max}}{g_{min}} (|\set{V}|-1) \right)
   \label{eq:gap}
\end{align}
bits of the capacity. The above result generalizes to multi-antenna nodes as well. 



\end{itwpaper}



\end{document}